# Comments on Coulomb pairing in aromatic hydrocarbons

D. L. Huber

Physics Department, University of Wisconsin-Madison, Madison, WI 53706

Abstract

Recently reported anomalies in the double-photonionization spectra of aromatic molecules such as benzene, naphthalene, anthracene and coronene are attributed to Coulomb-pair resonances of $\pi$ electrons.

E-mail address: huber@src.wisc.edu

1. Introduction

Recent studies of double-photoionization in aromatic hydrocarbons have revealed the existence of anomalous peaks in the distribution of doubly charged parent ions in partially deuterated benzene, naphthalene, anthracene, and coronene [1], pentacene [2], azulene and phenanthrene [3], and pyrene [4]. In the case of coronene and pyrene, two peaks were observed, while measurements carried out on the pentagonal rings pyrrole, furan, and selenophene [2] and thiophene [5] do not show a peak. In these papers, it was suggested that the anomalous peaks might be evidence for the existence of bound pairs of electrons analogous to the Cooper pairs associated with superconductivity [6,7].

The purpose of this note is to outline an interpretation of the anomalous peaks based on the existence of two-body bound states of interacting particles in a periodic potential. It was shown some time ago by Slater *et al* [8] and Hubbard [9] that such states exist for screened Coulomb interactions between electrons in one-dimensional metals with periodic potentials and short range repulsive interactions. In [10], an analysis is made of the two-electron bound states in a nearest-neighbor tight-binding chain with on-site and nearest-neighbor interactions, while a general analysis of the formation of bound states for short range repulsive forces is presented in [11]. It should be noted that the authors of [11], who refer to the bound states as 'Coulomb pairs, presciently pointed out the relevance of their calculations to the analysis of this note in the statement "Perhaps the simplest manifestation of the phenomenon under discussion is provided by the delocalized motions of paired electrons in a benzene ring structure.".

Our analysis is a straightforward application of the results of [10,11] to aromatic molecules. The starting point is a review of the HOMO states of a planar molecule with $N$-fold point symmetry. We use the Hückel approximation for the molecular orbitals neglecting overlap and considering only nearest-neighbor transfer. In this approximation, the energy levels are expressed as

$$E_v = -2\beta \cos(2\pi v / N) , \qquad (1)$$

where $\beta$ ($> 0$) is the transfer integral and $v = 0, \pm 1, \pm 2, \ldots, \pm N/2$ for $N$ even and $v = 0, \pm 1, \pm 2, \ldots, \pm(N-1)/2$ for $N$ odd. When $N$ is even there are two singlet states ($v = 0$ and $v = N/2$) and $N/2 - 1$ two-fold degenerate doublet states resulting in $N$ states overall; when $N$ is odd, there is one singlet ($v = 0$) and $(N-1)/2$ doublets, again with $N$ states overall. The wave functions corresponding to these energy levels can be expressed as complex amplitudes labeled by the eigenvalue index $v$ and the site label $n$. They take the form

$$\psi_v^n = N^{-1/2} \exp(i 2\pi v n / N), \qquad (2)$$

where $n = 1, 2, \ldots, N$. Note that in the case of the doublets, the wave functions come in pairs, corresponding to counter-propagating waves. When electron spin is taken into account, each of the doublets can be occupied by 4 electrons  Applying these arguments to the ground state of the simplest of the aromatics, benzene, where there are 6 carbon atoms and 6 $\pi$-orbitals, one finds the lowest level is an orbital singlet ($v = 0$), occupied by 2 electrons, followed by an orbital doublet ($v = 1$) occupied by 4 electrons; the $v = 2$ doublet and the $v = 3$ singlet are unoccupied.

2. Bound states

In the analysis of the Coulomb bound states we make use of the tight-binding results obtained in [10,12]. In the tight-binding approximation, which is appropriate when the basis states are Hückel orbitals, the Hamiltonian involves the transfer integral and two additional parameters associated with the Coulomb interaction: the Hubbard contact parameter [9], $U$, which we take to be positive corresponding to on-site Coulomb repulsion, and the nearest-neighbor Coulomb parameter, $J$, which we will also take to be positive. We further assume $U > J$. We focus on bound states where the center-of-mass momentum is zero. Because the bound state is comprised of two electrons, there are two types of spin states that are consistent with the overall antisymmetry of the two-particle wave function: the singlet spin state, $S = 0$, with a symmetric spatial wave function and the triplet spin states, $S = 1$, with an antisymmetric spatial wave function.

When $U \gg J$, the energy of the $S = 0$ two-electron bound state relative to twice the energy of the HOMO state is expressed as

$$E_B^{S=0} = (U^2 + 16\beta^2)^{1/2}, \qquad (3)$$

In the case of the $S = 1$ bound state, the $U$ parameter does not enter since the wave function vanishes when the particles occupy the same site. The bound state energy for this system is expressed as

$$E_B^{S=1} = J + (4\beta^2 / J). \qquad (4)$$

As pointed out in [10], the paired and free (Hückel) states overlap when $4\beta > J$. Also, since $U \gg J$, we have the inequality

$$E_B^{S=0} \gg E_B^{S=1}. \qquad (5)$$

3. Double-photoionization in benzene

As shown in [1-4], double-photoionization measurements carried out on partially deuterated benzene reveal a peak in the ratio of doubly charged parent ions to singly charged ions (after subtracting the contribution from the knock-out mechanism) vs photon energy located approximately 50 eV above the double ionization threshold. We argue the peak is related to the formation of a Coulomb pair comprised of electrons from the $\pi$-orbitals mentioned previously. The energy, 50 eV, is consistent with a $S = 0$ bound state. Given the value of $\beta$, we can use

Eq.(3) with $E_{HOMO} = -\beta$ to obtain the value of $U$. In the case of benzene, the HOMO-LUMO gap is equal to $2\beta$. Expressing the gap as the difference between the one-electron ionization energy and the electron affinity [13] leads to $\beta = 5.2$ eV. Setting $E_B^{S=0} + 2\beta = 50$ eV, we obtain $U = 45$ eV.

The tight-binding approach of Ref. [10] also provides an estimate of the exponential decay rate of the bound state, $\gamma$, which takes the form

$$\gamma = \text{arcsinh}(U/4\beta) = 1.51, \qquad (6)$$

using parameters appropriate to benzene. The exponential decay rate is defined terms of the decrease in amplitude of the localized state wave function at a site which has $m$ intervening bonds between it and the site of maximum amplitude. Thus if the maximum amplitude is $A$, the amplitude at the site three bonds away ($m = 3$) would be $\exp[-3 \times 1.51]A = 0.011A$.

4. Discussion

The 50 eV anomaly has been reported for deuterated benzene, coronene, naphthalene, anthracene, pentacene and pyrene, along with the isomers azulene and phenanthrene. It is likely that anomaly is associated with a $S = 0$ bound state. As mentioned, coronene and pyrene show a second peak at about 10 eV. It is possible that this peak is associated with a $S = 1$ bound state where the energy is given by Eq. (4). The absence of the low energy peak in the other compounds where there is a 50 eV peak may be due to the overlap of the $S = 1$ bound state with the Hückel states. The absence of a peak structure altogether in pyrrole, furan, selenophene and thiophene is likely caused by the presence of a non-carbon atom in the pentagonal ring interrupting the periodicity. A common feature of the compounds where the 50 eV peak has been detected is the presence of benzene-like hexagonal units which may indicate that the bound state is largely confined to a primary site and its near-neighbors as in benzene. The presence of the anomaly in the highly asymmetric azulene, where there are only singlet orbitals, suggests that electrons in both singlet and doublet orbitals can play a role in the higher symmetry molecules. It is likely that contributions from different orbitals are the primary cause of the width of the anomalous peak which is about 10 eV (HWHM, low energy side) [1-4] and the two-peak structure seen in coronene [4]. Finally, it must be emphasized that the rigorous analysis of the Coulomb pairing applies only to two-electron systems. For this reason, it may be more appropriate to speak of a Coulomb-pair resonance rather than a true bound state since the pair state rapidly decays into two free electrons.

Acknowledgment

The author would like to thank Ralf Wehlitz for preprints, references and many helpful comments.